# On the contribution of thermal excitation to the total 630.0 nm emissions in the northern cusp ionosphere

Norah Kaggwa Kwagala[1,2], Kjellmar Oksavik[1,2], Dag A. Lorentzen[2], and Magnar G. Johnsen[3]

[1]Birkeland Centre for Space Science, University of Bergen, Bergen, Norway, [2]University Centre in Svalbard, Longyearbyen, Norway, [3]Tromsø Geophysical Observatory, Arctic University of Norway, Tromsø, Norway

**Abstract** Direct impact excitation by precipitating electrons is believed to be the main source of 630.0 nm emissions in the cusp ionosphere. However, this paper investigates a different source, 630.0 emissions caused by thermally excited atomic oxygen O($^1$D) when high electron temperature prevail in the cusp. On 22 January 2012 and 14 January 2013, the European Incoherent Scatter Scientific Association (EISCAT) radar on Svalbard measured electron temperature enhancements exceeding 3000 K near magnetic noon in the cusp ionosphere over Svalbard. The electron temperature enhancements corresponded to electron density enhancements exceeding $10^{11}$ m$^{-3}$ accompanied by intense 630.0 nm emissions in a field of view common to both the EISCAT Svalbard radar and a meridian scanning photometer. This offered an excellent opportunity to investigate the role of thermally excited O($^1$D) 630.0 nm emissions in the cusp ionosphere. The thermal component was derived from the EISCAT Radar measurements and compared with optical data. For both events the calculated thermal component had a correlation coefficient greater than 0.8 to the total observed 630.0 nm intensity which contains both thermal and particle impact components. Despite fairly constant solar wind, the calculated thermal component intensity fluctuated possibly due to dayside transients in the aurora.

## 1. Introduction

The prominent dayside emission in the cusp ionosphere is the 630.0 nm resulting from the O($^1$D) state of atomic oxygen [*Rees et al.*, 1967; *Cogger et al.*, 1980]. The cusps are the regions where the solar wind has direct entry into the Earth's ionosphere. The highest probability of observing low-altitude cusp-like particle precipitation [e.g., *Newell and Meng*, 1988, 1992, 1994] and dayside cusp aurora [*Sandholt et al.*, 1998a, 2004] is between 10:30–13:30 magnetic local time (MLT) and around 75–80° magnetic latitude (MLAT), which is the MLAT range for Svalbard. For this reason, our study will focus on 12 ± 2 MLT over Svalbard, which we define as the area of the cusp ionosphere. Direct impact excitation of the O($^1$D) by precipitating electrons is believed to be the main source of the 630.0 nm emissions in the cusp ionosphere [*Solomon et al.*, 1988; *Meier et al.*, 1989]. However, the O($^1$D) state can also be excited by energetic electrons from the tail of the thermal electron distribution when the electron temperature is sufficiently high [*Meier et al.*, 1989; *Lockwood et al.*, 1993; *Mantas*, 1994; *Carlson et al.*, 2013]. Under such conditions, the thermal excitation component is anticipated to contribute significantly to the 630.0 nm emission although it is not well documented in literature.

The ambient electron gas in the ionosphere is believed to have a thermal or Maxwellian energy distribution where the population decreases exponentially with increasing energy. Therefore, the number of electrons above a particular energy is greatly dependent on the electron temperature. The O($^1$D) state has a low excitation threshold of 1.96 eV. If the electron temperature exceeds 3000 K, there may be enough electrons in the high-energy tail to excite observable emissions from the O($^1$D) state. At an electron temperature of 2000 K only 0.01% of a Maxwellian electron population has sufficient energy to excite the O($^1$D) state. At an electron temperature of 4000 and 6000 K this fraction increases to 1.5% and 7.3%, respectively [*Mantas*, 1994].

The 630.0 nm emission in the cusp is the result of soft (hundreds of eV) electron precipitation into the ionosphere [*Eather et al.*, 1979]. When a soft electron flux is incident in the atmosphere, half of the incident energy is carried out of the atmosphere [*Schunk and Nagy*, 1978]. The remaining energy is used to ionize and excite the neutral constituents mainly atomic oxygen in the F region and to heat the ambient electron gas by Coulomb collisions. The electron temperature can rise well above the ion and neutral temperatures to the point when the collision with ions can no longer effectively cool the electron gas. When the electron density is high, the electron gas cools by collision with ions, and thermal balance is attained at much lower electron temperature [*Lockwood et al.*, 1993]. This is because the rate of cooling to the ions is proportional to electron density $N_e$ times atomic oxygen density $N_o$ and $N_o = N_e$ in the F layer. Therefore, the rate of cooling to the ions is proportional $(N_e)^2$. The electron temperature is expected to be low and in good thermal contact when the electron density is enhanced. A combination of the electron temperature exceeding 3000 K and an electron density enhancement would indicate poor thermal balance. In such cases another cooling mechanism, collisional excitation of the O($^1$D) state, which can lead to 630.0 nm emissions, becomes important [*Carlson et al.*, 2013].

*Wickwar and Kofman* [1984] carried out a similar study to ours; however, their measurements were during sunlit conditions. In our study, we overcome this by using measurements from Svalbard during winter where we can make simultaneous optical measurements to also verify their results.

It is also important to note that echoes from non-Maxwellian distributions have been reported in incoherent scatter radar experiments at high latitude. Such events are referred to as naturally enhanced ion acoustic lines (NEIALs) [*Foster et al.*, 1988; *Michell et al.*, 2009; *Michell and Samara*, 2013; *Kontar and Pėcseli*, 2005; *Strømme et al.*, 2005]. NEIALs are captured as echoes which enhance one or both ion line shoulders in the typical power spectrum by 1–2 orders of magnitude [*Rietveld et al.*, 1991; *Collis et al.*, 1991], and enhancements up to 5 orders of magnitude have been reported as well [*Schlatter et al.*, 2013]. NEIALs have been observed in field-aligned measurements of incoherent scatter radars (ISRs). *Schlatter et al.* [2015] reported the NEIAL echoes to occur on field lines with particle precipitation. NEIALs have been observed in the F and even E regions during magnetically disturbed conditions [*Foster et al.*, 1988; *Rietveld et al.*, 1991; *Ogawa et al.*, 2006] with very high electron temperatures and 630.0 nm emission intensities at levels roughly above 15 kR [*Lunde et al.*, 2007; *Collis et al.*, 1991]. Standard analysis of ISR data, Grand Unified Incoherent Scatter Design and Analysis Package (GUISDAP) [*Lehtinen and Huuskonen*, 1996] assumes a Maxwellian distribution of the plasma. Consequently, the ISR parameters derived during NEIALs are unreliable.

In this paper we investigate the contribution of thermal excitation to the 630.0 nm emission in the cusp ionosphere. We derive the 630.0 nm emission rate due to thermal excitation using incoherent scatter radar measurements from Svalbard and compare it with optical observations from the same location. The instrumentation is described in section 2, and the observations are given in section 3. The discussion and conclusions are found in sections 4 and 5.

## 2. Instrumentation
### 2.1. The Meridian Scanning Photometer
The meridian scanning photometer (MSP) at the Kjell Henriksen Observatory (KHO) at Longyearbyen, Svalbard, is the main optical instrument used in this study. The KHO is located at 78.148°N, 16.043°E, and 520 m altitude. The MSP has five detector channels, where each channel consists of a narrow optical band-pass filter mounted onto a tilting frame. The tilting filter photometers are placed in front of a rotating mirror which scans the sky from north to south along the magnetic meridian and delivers intensity as a function of elevation angle in the meridian plane. The field of view is approximately one angular degree, and it takes 16 s to assemble one meridian scan with a typical spectral resolution of 0.4 nm. The five wavelengths include 630.0 nm, 427.8 nm, 557.7 nm, 486.1 nm, and 844.6 nm. For this study we mainly use the 630.0 nm channel which is the red line from the O($^1$D) excited state of atomic oxygen. The background is obtained by tilting each filter away from the peak emissions to an angle that transmits wavelengths representing the background emissions. The method of tilting enables the subtraction of the background from the peak emission [*Romick*, 1976]. Data from the MSP are absolutely calibrated in units of Rayleigh (R) which enables us to directly compare our calculations with observations. For studies of the dayside aurora, the MSP requires clear dark periods, which for Svalbard occurs in the winter months from late November to late January.

### 2.2. EISCAT Svalbard Radar
The EISCAT Svalbard Radar (ESR) is located 12 km southeast of Longyearbyen, Svalbard, at 78.15°N, 16.02°E, and 445 m altitude about 600 m north of KHO. The ESR operates in the 500 MHz band with a peak transmitted power of 1.0 MW. It has two dish antennas: a fully steerable parabolic 32 m dish and a fixed field-aligned 42 m dish. The ESR is ideal for studies of the cusp and polar cap. The MSP at KHO scans through the field of view of the 42 m dish at 81.6° elevation from the south. The geographic location of the ESR and MSP are shown in Figure 1. The black line shows the scanning field of view of the MSP, and the red circle marks the ESR 42 m beam which is colocated with the MSP. We searched the entire ESR 42 m data archive for periods when both the electron density and electron temperatures were enhanced with the electron temperature exceeding 3000 K. Two events, 22 January 2012 and 14 January 2013, satisfied the preceding conditions with available optical data.

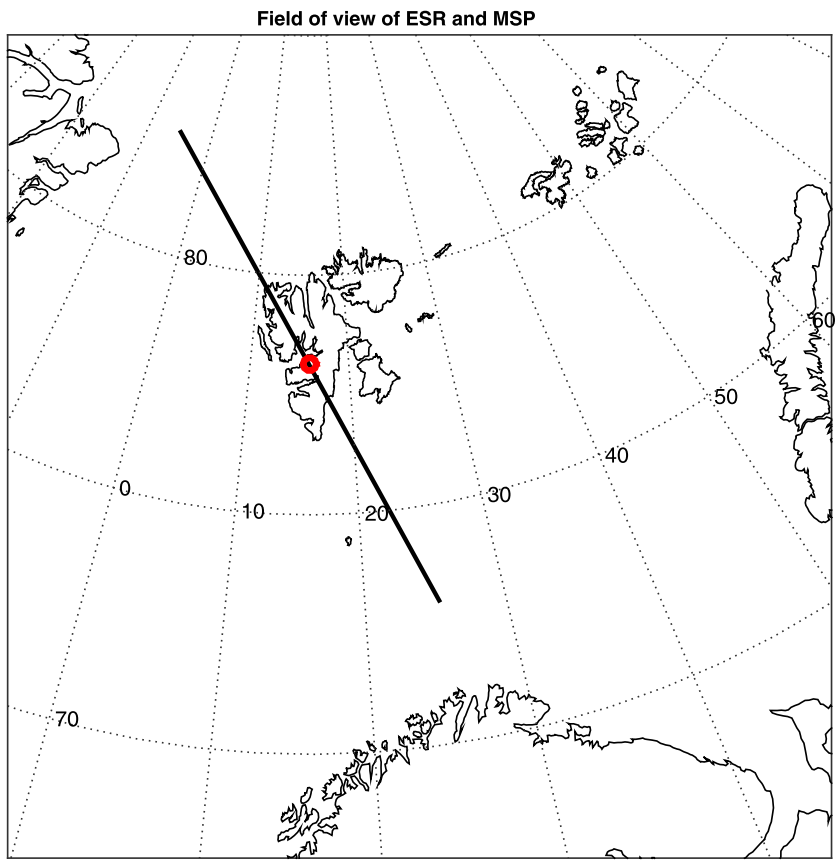

**Figure 1.** The black line indicates the field of view of the MSP, and the red circle indicates the ESR 42 m beam over Svalbard.

### 2.3. Solar Wind Data
Solar wind and interplanetary magnetic field (IMF) parameters were obtained from the GSFC/SPDF OMNI web interface. We used the high-resolution 5 min data which have already been time shifted to the magnetopause. The data include the 5 min averages of IMF components in GSM coordinates, solar wind speed, and proton density.

### 2.4. The Empirical Atmosphere Model
The density of atomic oxygen was generated from the empirical atmospheric model called the Naval Research Laboratory Mass Spectrometer and Incoherent Scatter Radar 2000 model (NRLMSISE-00) [*Picone et al.*, 2002].

### 2.5. Numerical Model: Calculation of Thermal Emission Rates
*Mantas and Carlson* [1991] reassessed the O($^1$D) thermal electron excitation rate putting into consideration both theoretical cross-section calculations [*Lan et al.*, 1972; *Thomas and Nesbet*, 1975] and experimental measurements of the cross section [*Doering and Gulcicek*, 1989]. They came up with a simple parametric equation for the thermal electron impact excitation rate $\alpha$ as a function of the electron temperature $T_e$. The cross-section calculations by *Lan et al.* [1972] have been reported to be the most complete so far. Using the cross sections presented by *Lan et al.* [1972], *Mantas and Carlson* [1991] recommended the equation

$$\alpha(T_e) = 0.15 \times \sqrt{T_e} \times \frac{(8537 + T_e)}{(34191 + T_e)^3} \times e^{\left(\frac{-22756}{T_e}\right)} \quad (\text{cm}^3/\text{s}) \quad (1)$$

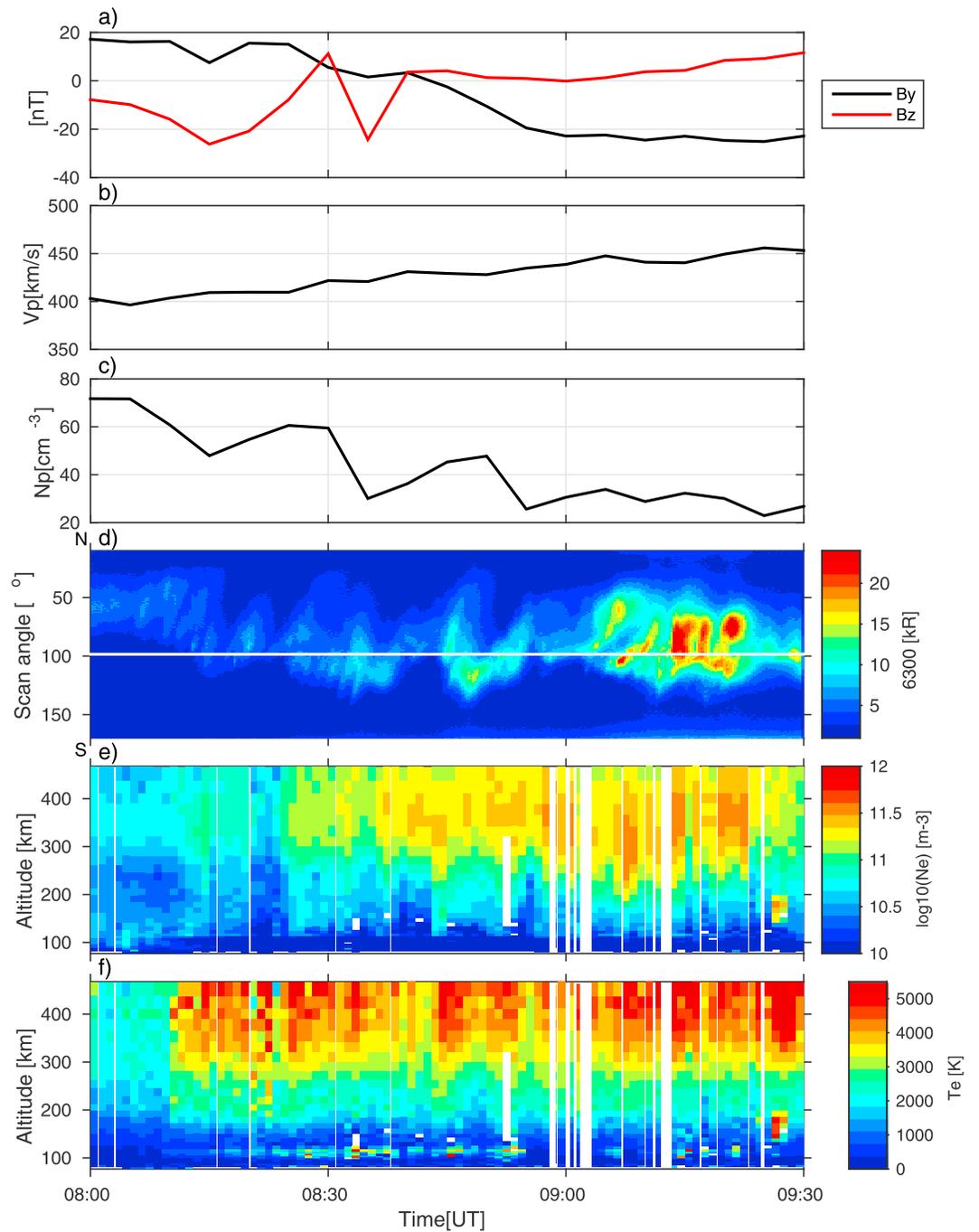

**Figure 2.** Solar wind and ionospheric data for Event 1 on 22 January 2012. (a) The interplanetary magnetic field components $B_y$ and $B_z$. (b) The solar wind speed. (c) The solar wind density. (d) Keogram from the meridian scanning photometer for the 630.0 nm emission line and the white line indicates the location of the ESR 42 m beam. (e) Electron density profiles. (f) Electron temperature profiles.

for the thermal electron impact $O(^1D)$ excitation rate $\alpha(T_e)$, where $T_e$ is the electron temperature in kelvin. The thermal excitation component depends on the electron density ($N_e$) and the number density of atomic oxygen ($N_o$) which are available to excite by collision, hence the excitation rate $\alpha(T_e)$. *Carlson et al.* [2013] employed equation (1) and derived the altitude ($h$) discriminated volume emission rate per kilometer (Rayleighs/km) of emission at 630.0 nm:

$$I_{630}(h) = \alpha\left[T_e(h)\right] \times N_o(h) \times N_e(h) \quad \text{(Rayleighs/km)} \quad (2)$$

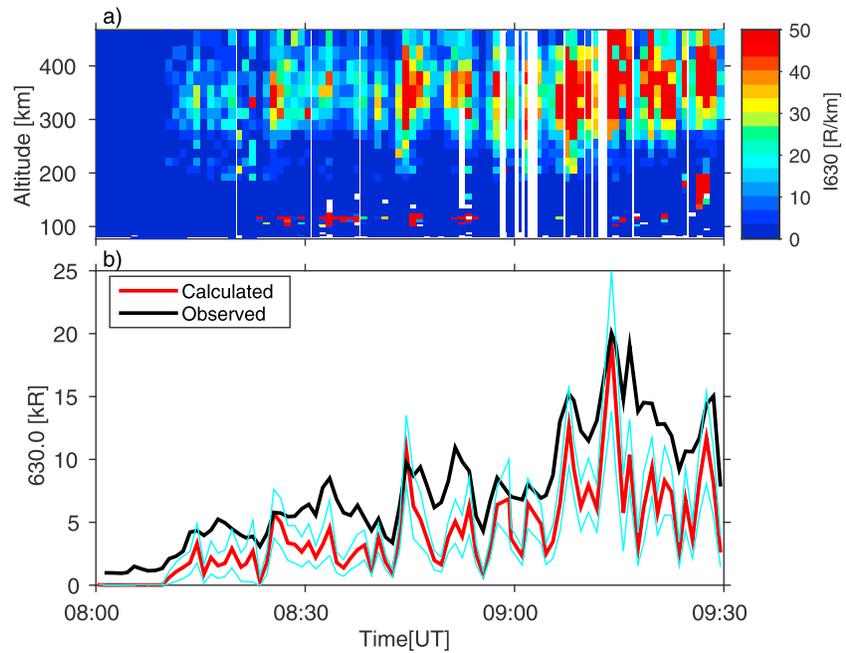

**Figure 3.** (a) The calculated volume emission rate versus height and (b) a comparison of the calculated thermal component with the MSP observations for event 1 on 22 January 2012. The cyan lines mark the upper and lower limits of the the error bars of the calculated thermal component.

which when integrated over the line of sight gives the total vertical column intensity for the emission at 630.0 nm:

$$I_{630} = \int_{250km}^{650km} I_{630}(h)dh \quad \text{(Rayleighs)} \qquad (3)$$

where $N_o(h)$ and $N_e(h)$ are measured in $cm^{-3}$, $T_e(h)$ in kelvin, and all variables are functions of altitude ($h$) in kilometers. For more details, see *Carlson et al.* [2013, and references therein].

## 3. Observations and Results
### 3.1. Event 1: 22 January 2012
Our first event is the interval 08:00–09:30 UT on 22 January 2012. Earlier on 22 January 2012 at 06:17 UT, the Earth was hit by a coronal mass ejection which led to very large interplanetary magnetic field and solar wind density compared to average values. Generally, this event is associated with disturbed conditions. In Figure 2a, the interplanetary magnetic field (IMF) $B_z$ component was southward at 08:00–08:28 UT. The interval 08:31–08:40 UT was dominated by a southward $B_z$, whereas the interval 08:40–09:30 UT was dominated by a negative $B_y$. The solar wind speed in Figure 2b gradually increased from 400 to 450 km/s throughout the interval of interest. In Figure 2c, there was a three-step decrease in the proton density from 70 $cm^{-3}$ at the beginning to 25 $cm^{-3}$ by the end of the interval of interest. We also observe that the solar wind parameters do not fluctuate much between 08:50 and 09:30 UT.

In the ionosphere, the ESR 42 m beam sampled 32 ranges at ∼77–470 km altitude along the magnetic field line, and the data were analyzed using a 60 s integration time. In Figure 2e and starting at 08:10 UT electron density enhancements exceeding $10^{11}$ $m^{-3}$ were observed at altitudes above 250 km throughout the period of interest. The electron density enhancement is accompanied by corresponding electron temperature enhancements (Figure 2f) exceeding 3000 K at altitudes above 300 km within the same time interval, as well as ion upflow (not shown). The white vertical regions are data gaps due to lack of data, bad fits, or NEIALs. The MSP (Figure 2d) observed 630.0 nm dominated emissions from 08:10 UT and throughout the period of interest. The 557.7 nm intensity (not shown) was around 3 kR, whereas the 630.0 nm intensity reached above 20 kR. The emissions migrated from higher-latitude characteristic of type 2 aurora to latitudes characteristic of type 1 aurora [*Sandholt et al.*, 1998a] where it remained in the ESR 42 m field of view (white line) from 08:10 to 09:30 UT. The strongest 630.0 nm emission was observed between 09:05 and 09:25 UT. There were no observations from the MSP after 09:30 UT due to twilight.

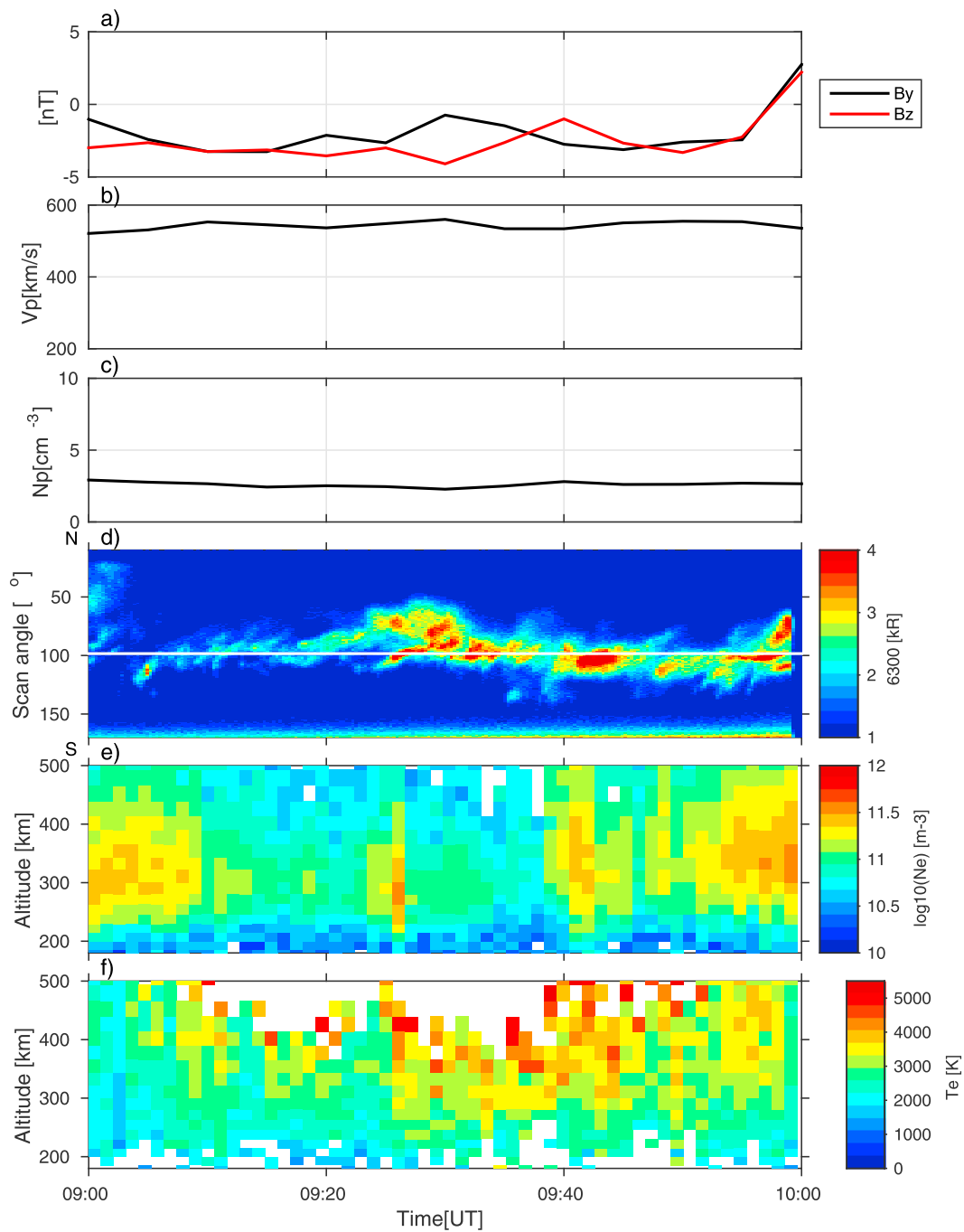

**Figure 4.** Solar wind and ionospheric data for event 2 on 14 January 2013. (a) The interplanetary magnetic field components $B_y$ and $B_z$. (b) The solar wind speed. (c) The solar wind density. (d) Keogram from the meridian scanning photometer for the 630.0 nm emission line, and the white line indicates the location of the ESR 42 m beam. (e) Electron density profiles. (f) Electron temperature profiles.

Figure 3a displays the thermally excited O($^1$D) 630.0 nm volume emission rate in Rayleighs per kilometer (R/km) calculated using equation (2). The 630.0 nm volume emission rate intensifies between 08:12 and 09:30 UT at altitudes above 250 km. The volume emission rate fluctuates possibly due to the horizontal motion of the aurora as a result magnetic reconnection events in the cusp [*Lockwood and Carlson*, 1992]. Using equation (3), we calculated the total column emission intensity resulting from the thermal excitation. Figure 3b shows the calculated thermal component and the total observed intensity from the MSP. After 08:15 UT, the thermal component on average contributes more than 50% of the total observed intensity by the MSP, which exceeds 10 kR at times.

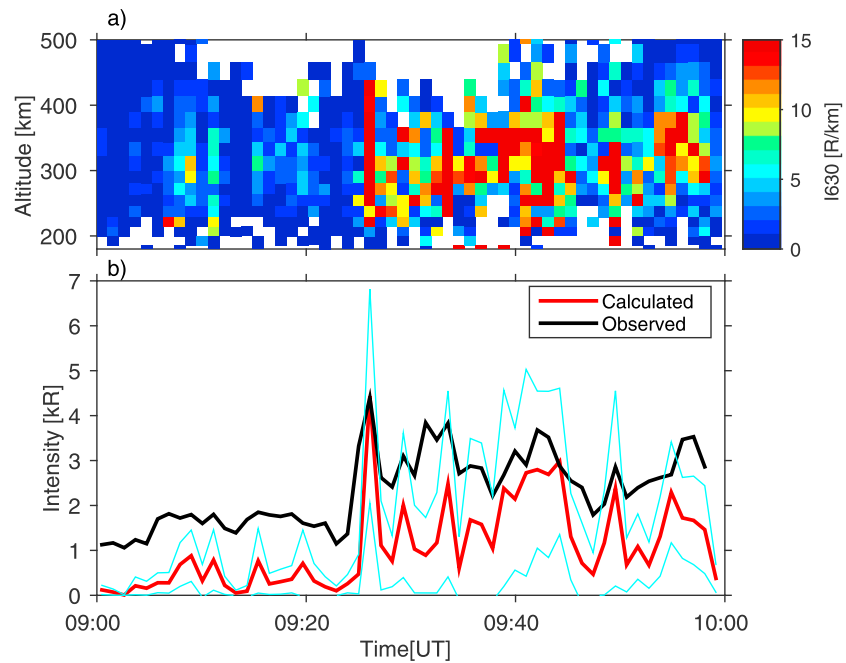

**Figure 5.** (a) The calculated volume emission rate versus height and (b) a comparison of the calculated thermal component with the MSP observations for event 2 on 14 January 2013. The cyan lines mark the upper and lower limits of the the error bars of the calculated thermal component.

### 3.2. Event 2: 14 January 2013

Our second event is 14 January 2013 between 09:00 and 10:00 UT. The solar wind and IMF conditions were more quiet. With regard to the solar wind, it is more of an average situation. Event 2 was characterized by a weak interplanetary magnetic field, high solar wind speed but low solar wind density. From Figure 4a, we see that $B_z$ and $B_y$ were ∼−3 nT most of the time. The solar wind speed (Figure 4b) is fairy stable between 520 and 560 km/s, while the proton density (Figure 4c) is ∼2.5 cm$^{-3}$.

In the ionosphere, the ESR recorded enhanced electron density (Figure 4e) exceeding $10^{11}$ m$^{-3}$ accompanied by corresponding electron heating to a temperature exceeding 3000 K (Figure 4f) and ion upflow (not shown) at altitudes above 350 km. The white vertical regions are data gaps. Figure 4d shows the MSP keogram for the 630.0 nm emission, and we see that the emission remained in the field of view of the radar most of the time. The observed aurora was dominated by the 630.0 nm emission, which is an indication of soft electron precipitation and heating of the ambient electron gas. Higher intensity was observed after 09:25 UT. Figure 5a displays the calculated volume emission rate for the thermally excited 630.0 nm emission. Figure 5b presents the comparison of the calculated total column intensity with the MSP observations. Generally, we see that the thermal component was on average ∼2 kR after 09:25 UT which contributed to around 50% of the total observed 630.0 nm emission by the MSP in 200–400 km altitude.

### 4. Discussion

This study has been limited to 630 nm emissions resulting from excitation of the O($^1$D) state by the heated ambient electron gas. The other important source of the 630.0 nm emission in the cusp is direct impact excitation of the O($^1$D) state from particle precipitation. We also calculated the estimated contribution from dissociative recombination (not shown) which was negligible during the period of interest.

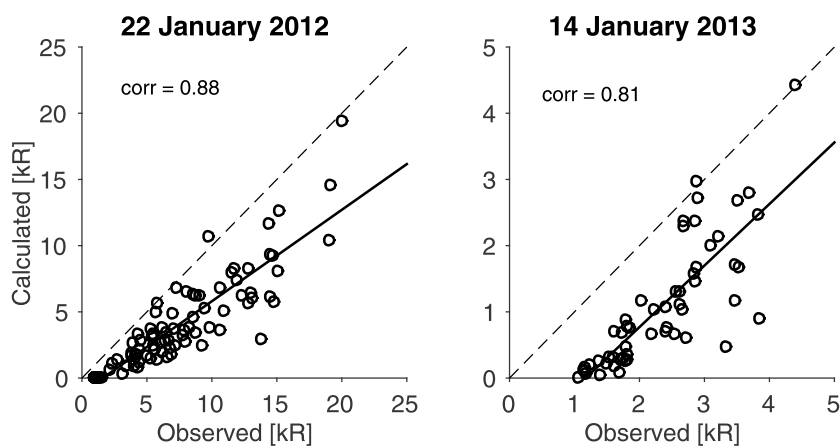

**Figure 6.** The calculated thermal component versus the total observed 630.0 nm MSP intensity on 22 January 2012 and 14 January 2013. The dashed line shows the one-to-one gradient line.

*Wickwar and Kofman* [1984] calculated the thermal component when the cusp was over Sondrestrom, and they measured high electron temperature. However, they did not have optical observations. *Carlson et al.* [2013] presented the first direct detection of the thermal component by applying the formula to ESR scan data and coincident all-sky images for the same latitudes as *Wickwar and Kofman* [1984]. We note that they used boundary tracking of the emission rate per kilometer altitude and not the total intensity. The scanning experiments of *Carlson et al.* [2013] are associated with large statistical error bars. In our study, we have used measurements from the fixed field-aligned 42 m ESR dish, which allows the data to be integrated in time in order to reduce statistical error bars at the expense of observation area. We made a general assumption that the thermal electron gas and the emitted thermal component remain in the ESR beam from excitation to emission of the 630.0 nm line. On this basis we have used the formula provided by *Carlson et al.* [2013] with smaller observational error bars, enabling us to better separate the thermal component from the total observed 630 nm intensities.

Figure 6 shows the correlation between the observed total emission and the calculated thermal component. The black line is a least squares fit to the compared intensities, with correlation coefficients of 0.88 and 0.81 for 22 January 2012 and 14 January 2013, respectively. The dotted line shows the slope of 1. This indicates that for both events the thermal component is a significant part of the observed 630.0 nm emission.

*Carlson et al.* [2013] emphasized magnetic reconnection as a major cause of electron heating in the cusp. During event 1, brief optical flashes were observed at 08:20:00, 08:25:30–08:26:00, and 08:32:00–08:35:00 UT in all-sky images from the University of Oslo camera at KHO (not shown). Unfortunately, these flashes were outside the field of view of the ESR and the MSP, so we could not test all the physical processes necessary for magnetic reconnection signatures [*Carlson*, 2012; *Carlson et al.*, 2004]. There were no optical flashes after 08.35:00 UT, which was mainly dominated by a strong $B_y$ component, $B_z \sim 0$ nT, and intense 630.0 nm emissions in the ESR beam. Using the same data set, *Ronksley* [2016] reported episodes of strong neutral upwelling after 08:25 UT to the end of our study period. Between 08:43 and 08:53 UT while $B_y$ changed from ~0 nT to ~−20 nT, the maximum vertical wind velocity of 200 ms$^{-1}$ was reported [*Ronksley*, 2016, Figure 7.2c] which coincides with one of the periods when the calculated thermal component contributed ~100% of the observed emission. The soft particle precipitation from the previously reconnected magnetic filed line may be contributing to the heating process, but another physical process may be participating as well. Low-latitude dayside magnetopause reconnection is not favored for the period between 09:05 and 09:25 UT, although the highest intensity from the thermal component is observed then. The ambient electrons are clearly heated by another physical process during this period. Lobe cell magnetic reconnection is a possibility under such IMF configuration [*Sandholt et al.*, 1998b], but we would expect it to occur at higher latitudes than the ones we are looking at. We also do not expect precipitation from lobe cell reconnection to heat the ambient electron gas to such a high temperature as measured during the time of consideration. The ambient electrons could also be heated by ions when there is intense frictional heating in the F region. However, this requires that the ions are hotter than the electrons, yet for the period of interest here, the electrons are hotter than the ions.

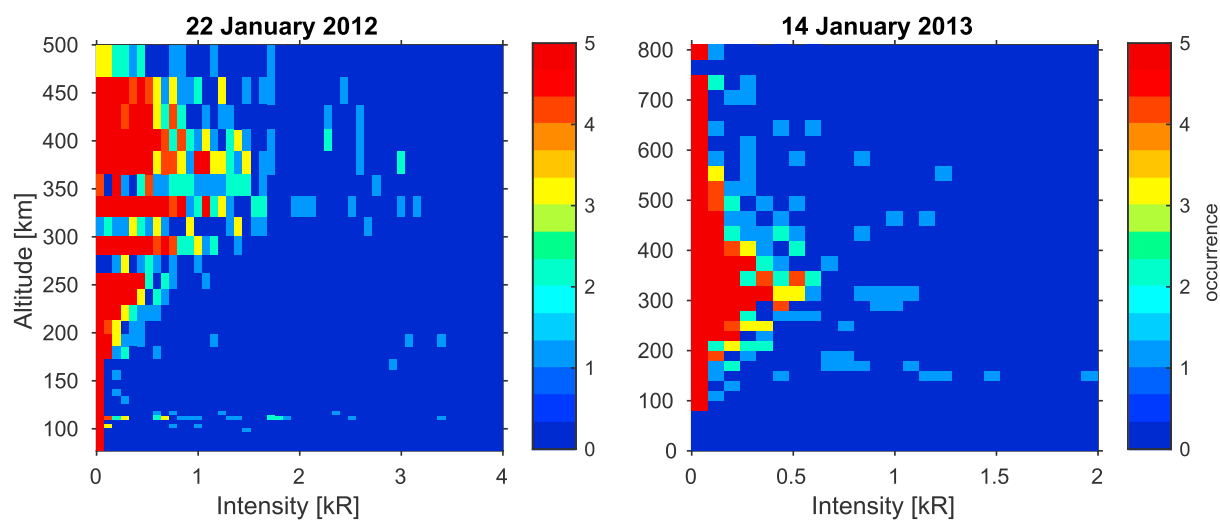

**Figure 7.** Occurrence of thermal emissions versus altitude and intensity for the events on 22 January 2012 and 14 January 2013.

Therefore, it is less likely for the ions to be the heating source of the ambient electrons. Given that event 1 is a highly disturbed period, as discussed above the specific heating process is not clear.

Event 2 had southward $B_z$ with some 630 nm flashes both in the MSP and the all-sky images. An outstanding example but weak signature is seen in Figure 4 between 09:23 and 09:29 UT. During this period, a poleward moving auroral form [e.g., *Fasel*, 1995; *Sandholt and Farrugia*, 1999; *Milan et al.*, 2000; *Sandholt et al.*, 2004; *Oksavik et al.*, 2004; 2005; *Carlson et al.*, 2006; *Moen et al.*, 2008; *Lorentzen et al.*, 2010] crosses the ESR beam at 09:25 UT seen in the 630.0 nm keogram, and at the same time the electron density is enhanced below 200 km with an electron temperature enhancement near and above 300 km, all of which are signatures highlighted by *Carlson et al.* [2004] but not isolated from the background structure. Therefore, this event has greater possibility to be driven by magnetic reconnection pulses [*Carlson et al.*, 2004], but no clear signatures were confirmed. We also note that for both events the magnitude of $B_y$ was greater than zero, and the greatest thermal component intensity was during event 1 on 22 January 2012 when $B_z$ was northward. This emphasizes the effectiveness of the IMF $B_y$ component in the high-latitude ionospheric processes as suggested in previous studies [e.g., *Knipp et al.*, 2011; *Li et al.*, 2011].

The peak emission altitude has an implication on the magnetic reconnection energy entering the cusp ionosphere, since the estimate of reconnection voltage varies as the square of the assumed emission altitude [*Lockwood et al.*, 1993]. An estimate of the altitude of the thermal component for both events is shown in Figure 7. Instead of integrating over the whole altitude of interest as in equation (3), we integrate over each range gate separately. Bins of intensity and altitudes are made. The occurrence of the resulting intensities at different altitudes is presented in Figure 7. We assume that the O($^1$D) below 250 km is greatly quenched. The altitude with the highest occurrence is used as the estimate for the peak emission altitude. The estimated peak emission altitude was 378 ± 35 km for event 1 and 330 ± 30 km for event 2. Theoretically, *Kozyra et al.* [1990] suggested that the emission altitude for the thermal component should be expected near 500 km which is higher than the estimated peak emission altitudes for the events studied here. However, the peak emission altitude for the event on 22 January 2012 is closer to 400–450 km reported by *Carlson et al.* [2013]. The estimated peak emission altitudes for both events studied in our work are in agreement with thermal component emission peak near 350 km reported by *Wickwar and Kofman* [1984]. As expected, the thermal component did not contribute to all the observed 630.0 nm emission, which implies that the total emission is due to a combination of different processes like direct impact excitation in addition to the thermal excitation. Hence, consistent with the fact that the precipitating soft electrons produce both thermal and impact emissions on the same flux tube, however, the peak emission altitude would differ [*Lockwood et al.*, 1993].

It is important to point out that the ESR 42 m data were checked for the whole dayside, but the thermal component only became significant when the cusp was over Svalbard. In the scanning experiment used by *Carlson et al.* [2013], thermally excited 630.0 nm emissions were also detected in the cusp sector. We note that we were limited to the winter period due to optical observations, but such events may in principal also exist in other seasons. We leave that for future work.

When we consider the errors associated with the measured electron temperature and electron density we get the upper and lower error boundaries in the calculated thermal component. These boundaries are indicated by the cyan lines in Figures 3 and 5. For event 1, the error boundaries do not deviate greatly from the final thermal component. However, for event 2 where the intensity is generally low the error boundaries are greater in comparison to the calculated thermal component. We have also assumed that the empirical atmospheric model produces reasonable results which is more uncertain for the disturbed ionospheric conditions [*Vickers et al.*, 2013]. A 20–30% underestimate of the atmospheric density at cusp latitudes from the empirical model was reported near solar maximum [*Liu and Lühr*, 2005; *Forbes et al.*, 2005]. Based on this, the contribution presented here is the lower limit. When we redo our analysis with an atomic oxygen density that is 30% higher than the model predictions, the resulting thermal excitation component will only be increased by 10–15%. Previous studies have also suggested that during large IMF $B_y$ conditions, the empirical models may be unable to respond and hence could potentially underestimate the neutral density by 100% [e.g., *Crowley et al.*, 2010]. This further emphasizes that the atomic oxygen density used herein is the lower limit.

## 5. Summary and Conclusion

We have presented the first direct comparison between calculated thermally excited and observed 630 nm intensities. For the two events studied, the thermal component was highly correlated (correlation coefficient ∼0.8) with the MSP observations with an average contribution of ∼ 50% for both a disturbed and a relatively quiet cusp. The specific source of the electron gas heating is not certain, but the cooling by thermal excitation appears clear.

In agreement with earlier studies, the thermal component was significant when the electron temperature exceeded 3000 K with electron density enhancements between $(1–5) \times 10^{11}$ m$^{-3}$ [*Egeland et al.*, 1992]. In agreement with earlier studies [*Carlson et al.*, 2013; *Johnsen et al.*, 2012; *Kozyra et al.*, 1990; *Egeland et al.*, 1992], our estimated emission altitude for the thermal component is higher than the generally accepted emission altitude of 225–250 km for the 630.0 nm emission [*Lockwood et al.*, 1993]. Our findings agree with *Egeland et al.* [1992] that the thermal excitation may be responsible for the 630.0 nm emission at 350 km and above.

**Acknowledgments**
The EISCAT data were accessed from https://www.eiscat.se and processed using GUISDAP. The Meridian Scanning Photometer (MSP) data at Kjell Henriksen Observatory (KHO) were provided by Dag Lorentzen. The interplanetary magnetic field and solar wind data were provided by the NASA OMNIWeb service (https:omniweb.gsfc.nasa.gov/). The NRLMSISE-00 Atmospheric model was accessed from https://ccmc.gsfc.nasa.gov/modelweb/models/nrlmsise00.php. This project has been funded by the Norwegian Research Council under contracts 223252 and 212014.